\documentclass[aps,pra,preprint,groupedaddress,showpacs,showkeys]{revtex4}
\usepackage{amssymb,bm}
\usepackage{graphicx}
\usepackage{amsmath}

\begin{document}

\title{Final state Coulomb interaction and asymmetry of pair production close to  threshold in $e^+e^-$ annihilation. }

\author{V.F. Dmitriev}\email{V.F.Dmitriev@inp.nsk.su}
\author{A. I. Milstein}\email{A.I.Milstein@inp.nsk.su}
\affiliation{Budker Institute of Nuclear Physics, 630090 Novosibirsk, Russia}

\date{\today}

\begin{abstract}
We investigate a contribution of the $d$ wave to the cross section of $e^+e^-$ annihilation to the pair of charged leptons or nucleons close to  threshold of the process.
In contrast to the point of view accepted in literature,   due to the Coulomb final state interaction this contribution does not vanish even at zero relative velocity of produced particles. This  results in the nonzero asymmetry in  angular distribution at threshold. Though value of the asymmetry is small, observation of this effect is not hopeless.
\end{abstract}
\maketitle

\section{Introduction}

Recently, much attention has been attracted to  low-energy nucleon-antinucleon production in the processes  $e^+e^- \to p\bar p$  and $e^+e^- \to n\bar n$.
It was  found that the ratio $|G_E(Q^2)/G_M(Q^2)|$ of the
proton (antiproton) electric, $G_E(Q^2)$, and  magnetic, $G_M(Q^2)$, form factors  strongly depends on $Q^2=4E^2$ (in the
center-of-mass frame) in the narrow  region of  the energy $E$  near  threshold of $p\bar p$ production  \cite{Bardin94,Armstrong93,Aubert06}.
The most natural explanation of this behavior refers to the  final state interaction (FSI) of  proton and antiproton
\cite{Ker04,Bugg04,Zou04,Loi05,Sib05,Sib06, DM2007,BDMS2011}. Such behavior of the form factors results in the  strong energy dependence of the angular distribution  of  final particles near  threshold. Similarly, angular distribution   of heavy quarks and leptons produced in $e^+e^-$ annihilation close to threshold depends strongly  on the energy \cite{FK1987, BHKT1995, HT1998, MY1998}.

It is interesting to consider  influence of the Coulomb FSI on energy and angular dependence of the cross section  of particle production in
  $e^+e^-$ annihilation. This is so-called Sommerfeld-Gamov-Sakharov factor \cite{SGS} discussed in many
   papers \cite{Byers1990,Kaiser2003,Vol2005,Baldini2009,Iengo2009,Cassel2010}. The origin  of this factor is simple.
Let us consider the attractive Coulomb potential  $U(r)=-e^2/r$ and the corresponding radial   wave function   $R_{k\, l}^{(c)}(r)$, where $e$
is the electric charge, $k$ is the particle momentum,
 and $l$ is the angular momentum. Then the ratio $C_{k\,l}= |R_{k\, l}^{(c)}(r)/R_{k\, l}^{(0)}(r)|^2$ at $kr\longrightarrow 0$, where  $R_{k\, l}^{(0)}(r)$
 is the radial wave function for zero  charge,  has the form \cite{LL1977},
\begin{eqnarray}\label{C}
C_{k\,0}=\frac{2\pi\eta}{1-e^{-2\pi \eta}}\,,\quad C_{k\,l\ne 0}=C_{k\,0}\prod_{s=1}^l\left(1+\frac{\eta^2}{s^2}\right)\,,\quad\eta=\frac{\mu \alpha}{k}\,.
\end{eqnarray}
Here $\mu$ is the reduced mass of the system, $\alpha=e^2$ is the fine structure constant,  units $\hbar=c=1$ are used. If pair of particles is produced with the angular momentum
 $l$ and energy close to the threshold, then the cross section $d\sigma/d\Omega$ is equal to the cross section $d\sigma_B/d\Omega$,
  calculated in the leading Born approximation,  multiplied by the factor $C_{k\,l}$ \cite{Iengo2009,Cassel2010,BZP1966}
  with the replacement $\mu\longrightarrow M/2$ ($M$ is the mass of the particle ), and with $k$ being the momentum of each particle in the center-of-mass frame.
   At $k\longrightarrow 0$,  the cross section $d\sigma_B/d\Omega$ is proportional to $k^{2l+1}$. For $\eta\gg 1$, we have
   $C_{k\,l}=2\pi \eta^{2l+1}/(l!)^2\propto 1/k^{2l+1}$, so that   $d\sigma/d\Omega$  is nonzero constant at $k=0$ for all $l$.
In  $e^+e^-$ annihilation via one virtual photon to  pair of particles with spin $1/2$, the only contribution of partial waves with the angular momentum $l=0$ ($s$ wave) and $l=2$ ($d$ wave) exists. It is generally accepted   in literature that  $d\sigma/d\Omega$ is  angular independent at $k=0$ since the contribution of $d$ wave vanishes. However, the  arguments presented above show that  the contribution of $d$ wave may also be nonzero at $k=0$ due to Coulomb interaction, so that  $d\sigma/d\Omega$ is not spherically symmetric at threshold. This effect is interesting both from the theoretical and experimental point of view, and we investigate it in detail in the present paper.

\section{Angular dependence of the cross-section}

Let us discuss first the production of a lepton pair.
In nonrelativistic approximation, the Born amplitude of the process  $e^+e^-\longrightarrow L\bar{L}$ near  threshold
via one virtual photon  can be presented as follows (in units $4\pi\alpha/Q^2$) \cite{BLP1971} :
\begin{eqnarray}\label{1}
T_{\lambda\mu}^B =\sqrt{2} \bm{\epsilon}_\lambda^*\cdot \Bigg[G_s
\bm e_\mu  +G_d\frac{{\bf k}^2{\bf e}_\mu-3({\bf k}\cdot{\bf
e}_\mu){\bf k}}{6M^2}\Bigg]\, .
\end{eqnarray}
Here ${\bf e}_\mu$ is a virtual photon polarization vector,
$\bm{\epsilon}_\lambda$ is the spin-1 function of $L\bar{L}$ pair, $G_s=F_1(Q^2)+F_2(Q^2)$,
$G_d=F_1(Q^2)-F_2(Q^2)$, $F_1(Q^2)$ and $F_2(Q^2)$ are the Dirac form factors of lepton.
Two tensor structures in Eq.(\ref{1}) correspond to the s-wave and
d-wave production amplitudes. The total angular momentum of the
$L\bar{L}$ pair is fixed by a production mechanism.
 We skipped the correction of order $k^2/M^2$ in the contribution of $s$ wave since it does not contribute to asymmetry. Note that   the Dirac form factor  $F_2(Q^2)$  is small  ($\propto\alpha$) and the form factor $F_1(Q^2)$ is almost constant in the vicinity of threshold. The amplitude which takes into account  the effect of Coulomb interaction in the final state can be written as follows
\begin{eqnarray}\label{2}
T_{\lambda\mu} & =&\sqrt{2}\bm{\epsilon}_\lambda^* \cdot \int \frac{d^3p}{(2\pi)^3}{\Phi}^{(-)*}_{\bf k}({\bf p})
\Bigg[
G_s\bm e_\mu +G_d\frac{{\bf p}^2{\bf e}_\mu-3({\bf p}\cdot{\bf
e}_\mu){\bf p}}{6M^2}\Bigg]\, ,
\end{eqnarray}
where $\Phi^{(-)}_{\bf k}({\bf p})$ is the Fourier transform of the function $\psi^{(-)}_{\bf k}({\bf r})$, the wave function
of the $L\bar{L}$ pair in  coordinate space (see, e.~g., \S 136 and \S 137 of~\cite{LL1977}):
\begin{eqnarray}
\psi^{(-)}_{\bf k}({\bf r})&=&
\frac{1}{2k} \sum_{l=0}^\infty  \imath^l \, {\rm e}^{-{\rm
i}\delta_l}\, (2 l + 1)\, R_{k\,l}(r) \, P_l(\bm{n} \cdot
\hat{\bm k})\,,\nonumber\\
\delta_l&=&\mbox{arg}\Gamma(l+1-\imath\eta)\,.
\end{eqnarray}
Here $\bm n=\bm r/r$, $\hat{\bm k}=\bm k/k$, $P_l$ are the
Legendre polynomials, and $R_{k\,l}(r)$
is the regular radial solution of the
Schr\"{o}dinger equation in a Coulomb field,
\begin{eqnarray}
R_{k\,l}(r)=\frac{2k\,l!\,\sqrt{C_{kl}}}{(2l+1)!}(2kr)^l\,e^{-\imath kr}\,F(\imath \eta+l+1,2l+2,2\imath kr)\,,
\end{eqnarray}
where $C_{kl}$ and $\eta$ are defined in Eq.(\ref{C}), and $F(a, b, x)$ is a confluent hypergeometrical function. Performing the integration over $\bm p$ in Eq.(\ref{2}), we obtain
\begin{eqnarray}\label{3}
T_{\lambda\mu} & =&\sqrt{2}\bm{\epsilon}_\lambda^*\cdot \Bigg[
e^{i\delta_0}\sqrt{C_{k\,0}}G_s\,\bm e_\mu  + e^{i\delta_2}\sqrt{C_{k\,2}}G_d\,\frac{{\bf k}^2{\bf e}_\mu-3({\bf k}\cdot{\bf
e}_\mu){\bf k}}{6M^2}\Bigg]\nonumber\\
&=&\sqrt{2}\,e^{i\delta_0}\sqrt{C_{k\,0}}\,\bm{\epsilon}_\lambda^*\cdot \Bigg[G_s
\bm e_\mu  + \left(1-\frac{\eta^2}{2}-\frac{3i\eta}{2}\right)\,G_d\,\frac{{\bf k}^2{\bf e}_\mu-3({\bf k}\cdot{\bf e}_\mu){\bf k}}{6M^2}\Bigg]\,.
\end{eqnarray}
We see that second term in a square brackets is finite at $k\rightarrow 0$. Let us introduce the new "dressed" form factors $ {\cal G}_d$ and  $ {\cal G}_s$ via the relation
\begin{equation}\label{calg}
{\cal G}_d=\left(1-\frac{\eta^2}{2}-\frac{3\imath\eta}{2}\right)G_d \,,\quad   {\cal G}_s= G_s\,.
\end{equation}
Due to FSI, the form factor  $ {\cal G}_d$ becomes singular at $\eta \rightarrow \infty$ ( $k\rightarrow 0$). For  the corresponding "dressed" Dirac form factors, we have
\begin{eqnarray}\label{calf1f2}
&& {\cal F}_1= \left(1-\frac{\eta^2}{4}-\frac{3i\eta}{4}\right)F_1+\left(\frac{\eta^2}{4}+
\frac{3i\eta}{4}\right)F_2 \,,\nonumber\\
&&{\cal F}_2=\left(\frac{\eta^2}{4}+
\frac{3i\eta}{4}\right) F_1+\left(1-\frac{\eta^2}{4}-\frac{3i\eta}{4}\right)F_2 \,,
\end{eqnarray}
so that both ${\cal F}_1$ and ${\cal F}_2$ are singular at   $k\rightarrow 0$ (see also Ref.\cite{BZP1966}). Note that ${\cal F}_2$ is not equal to zero even in the case $F_2=0$.

  The cross section corresponding to the amplitude (\ref{3}) has the form in the center-of-mass frame
\begin{equation}\label{CS}
\frac{d\sigma}{d\Omega}=\frac{\beta\alpha^2}{2Q^2}C_{k\,0}\,\Bigg[|{\cal G}_s|^2+\frac{\beta^2}{3}\,\mbox{Re}({\cal G}_s^*{\cal G}_d)
P_2(\cos\theta)+\beta^4|{\cal G}_d|^2[2-P_2(\cos\theta)]\Bigg].
\end{equation}
where $\beta=k/M$,  and $\theta$ is the angle between the electron (positron) momentum $\bm P$ and the momentum of the final particle $\bm k$. Here we have performed  summation over the polarization of $L\bar L$ pair and averaging over polarization of virtual photon,
\begin{equation}
\sum_{\lambda=1,2,3} \epsilon_\lambda^{i*}\epsilon_\lambda^j=\delta^{ij}\,,\quad
 \frac{1}{2}\sum_{\mu=1,2} e_\mu^{i*}e_\mu^j=\frac{1}{2}\delta^{ij}_\perp=\frac{1}{2}(\delta^{ij}-P^iP^j/P^2)\,.
\end{equation}
Omitting the form factor $F_2(Q^2)$ for lepton production, we obtain from Eq.(\ref{CS})
\begin{equation} \label{CSL}
\frac{d\sigma}{d\Omega}=\frac{\beta\alpha^2}{2Q^2}C_{k\,0}|F_1(Q^2)|^2\,\Bigg[1+
\frac{1}{3}(\beta^2-\frac{\alpha^2}{8})\,
P_2(\cos\theta)\Bigg]\,
\end{equation}
where  $\eta=\alpha/(2\beta)$.
  Thus, at $\beta=0$ the coefficient in front of the Legendre  polynomial $P_2$ is $-\alpha^2/24$. The standard equation for the cross section expressed via the electric form factor $G_E$ and magnetic form factor $G_M$ reads (see, e.g., Ref. \cite{Baldini2009})
\begin{eqnarray}\label{CS0}
\frac{d\sigma}{d\Omega} & =&\frac{\beta\alpha^2}{4Q^2}C_{k\,0}\,\Bigg[|G_M(Q^2)|^2(1+\cos^2\theta)+\frac{4M^2}{Q^2}\,
|G_E(Q^2)|^2\sin^2\theta\Bigg]\,.
\end{eqnarray}
In terms of the "dressed" form factors ${\cal G}_s$ and ${\cal G}_d$ the electromagnetic Sachs form factors, which take into account Coulomb FSI, have the form
\begin{eqnarray}
G_M={\cal G}_s+\frac{\beta^2}{6}{\cal G}_d\,,\quad \frac{2M}{Q}G_E={\cal G}_s-\frac{\beta^2}{3}{\cal G}_d\,.
\end{eqnarray}
Since $\beta^2{\cal G}_d$ does not vanish in the limit $\beta\rightarrow 0$, we find that
$G_E\ne G_M$ at threshold, in contrast to the conventional statement. Unlike the form factors ${\cal G}_s$ and  ${\cal G}_d$, the electromagnetic Sachs form factors $G_E$ and $G_M$ are not singular in the limit  $\beta\rightarrow 0$.

For $p\bar{p}$ production near threshold, in addition to Coulomb interaction, there is a strong interaction that has to be accounted together with the Coulomb FSI. However, due to different scales in the strong and the Coulomb interactions, their contributions can be separated. The strong interaction exists at small distances, $r\leq 1.5$ fm, where the Coulomb interaction can be neglected. In the absence of the long range interaction all the effects of strong interaction are absorbed into two complex functions $F_1(Q^2)$ and $F_2(Q^2)$, the Dirac form factors of proton. Typical scale in the Coulomb wave function at zero kinetic energy is given by Bohr radius $a_B=2/(M_p\alpha)$, $a_B \sim 57$ fm. At small distances, $r < 1$ fm, the Coulomb wave function is a smooth function, and one can use Eq.(\ref{3}) for the amplitude,  where the s-wave and d-wave amplitudes depend on complex proton form factors $F_1$ and $F_2$. For $p\bar{p}$ production, the cross-section will be given  by Eqs.(\ref{calg}) and (\ref{CS}).

For $t\bar{t}$  quark-antiquark pair production in $e^+e^-$ annihilation, the angular asymmetry of the corresponding jets can be estimated by Eq.(\ref{CS0}) with the replacement
$\alpha \rightarrow \frac{4}{3}\alpha_s(q)$, where $\alpha_s(q)$ is the QCD running coupling constant at the point $q $. Strictly speaking, the threshold of this reaction is not fixed due to finite width of $t$-quark, $\Gamma_t \sim 1.4$ GeV. The width defines the minimal value of $q$ and the corresponding maximal value of $\alpha_s(q_{min})$: $q_{min}\sim \sqrt{M_t \Gamma_t}\sim 15.5$~GeV, and $\alpha_s(q_{min})\approx 0.2$, see Ref.\cite{FK1987}. Therefore, typical asymmetry in the vicinity of threshold is of the order of $\sim 2 \alpha_s^2(q_{min})/27 \sim 3\times 10^{-3}$.
Of course, to determine the axes of jet  at small $\beta$ is a separate problem.
\section{Conclusions}
We have calculated the asymmetry in angular distribution of pair production in $e^+e^-$ annihilation taking into account the Coulomb FSI. It is shown that the asymmetry does not vanish in the limit $\beta \rightarrow 0$. The origin of this anomaly is the singularity of the "dressed" Dirac form factors ${\cal F}_1$ and ${\cal F}_2$  at $\beta\rightarrow 0$, see Eq. (\ref{calf1f2}).
The corresponding electromagnetic Sachs form factors $G_E$ and $G_M$ are not singular, but $G_E \ne G_M$ at $\beta\rightarrow 0$. The effect is non-perturbative , since in this limit $\alpha/\beta \gg 1$. This very  nonzero difference $G_E-G_M$ at $\beta\rightarrow 0$ provides  nonzero asymmetry at threshold. Although the value of the asymmetry is small for electromagnetic Coulomb interaction ($\sim \alpha^2/24)$,  the asymmetry for heavy quark pair production, where the Coulomb-like strong interaction must be considered, can be noticeable.

\section*{Acknowledgements}
The work  was supported by the Ministry of Education and Science of the
Russian Federation and in part (VFD) by RFBR grant 12-02-00560\_a, DFG grant, GZ: HA 1457/7-2.

\end{document}